# Tissue equivalence of some phantom materials for proton beams

V.N. Vasiliev[1*], V.I. Kostjuchenko[2], O.B. Riazantsev[2], V.G. Khaybullin[2],
S.I. Samarin[3], A.S. Uglov[3]

[1]Institute for Nuclear Research, Russian Academy of Sciences, Moscow, Russia
[2]Institute for Theoretical and Experimental Physics, Moscow, Russia
[3]Russian Federal Nuclear Center–Zababakhin Institute of Applied Physics, Snezhinsk, Russia

**Abstract:** Tissue and water equivalence of some phantom materials originally developed for conventional radiation therapy was investigated on the ITEP medical proton beam facility. The proton CSDA range in three variants of Plastic Water, lung, adipose, muscle and compact bone substitute materials (CIRS Inc., USA) was measured by a silicon diode as well as the residual proton range in liquid water after passing a slab of each material under investigation. In addition, the proton range in five materials of known elemental composition was calculated by Monte Carlo technique. The obtained results were compared with reference data from ICRU report 49 for respective biological tissues and water. A total uncertainty of the proton range ratios was estimated to be from 0.9 to 1.5% (1SD). Within these uncertainties, Plastic Water, Plastic Water LR, Plastic Water DT, muscle and compact bone demonstrated a good agreement with the reference data. The range in adipose and lung substitutes is a few percents lower than that in the respective tissues.

## Introduction

Application of proton therapy in medical practice requires a special equipment to simulate a human body, organs and tissues as well as the reference medium, water. This equipment includes some dosimetry phantoms and test objects and has to be used both in beam commissioning and dose distributions verification process. The proton interaction cross sections of the phantom materials are to be close as possible to those of respective biological tissues. This problem is well known in conventional radiation therapy by photon and electron beams. Numerous tissue and water equivalent materials were developed for radiation therapy last decades [1-3] and high accuracy in the interaction cross sections simulation was achieved for modern substitutes. In particular, water and tissue equivalent plastics manufactured by CIRS Inc. are close to liquid water and respective tissues within 0.5-1.0% for photon and electron beams. Nevertheless, the applicability of these materials in proton beams is to be specially verified.

In this work, we have performed an experimental and theoretical evaluation of tissue and water equivalence of seven phantom materials originally developed for conventional radiation therapy. A proton range ratio measured in each material under investigation and in liquid water was used as an equivalence estimator and compared with theoretical values for respective tissues provided by ICRU Report 49. In addition, proton ranges in four plastic samples of known elemental composition were simulated by Monte Carlo technique.

The results of the comparison allowed evaluating the equivalence of these plastics respective to proton range and stopping power. Other significant parameters, the scattering power and the inelastic nuclear reaction contribution, are planned for investigation in the future.

## Materials and methods

Three types of Plastic Water (developed for high, medium and low photon energy), adipose, muscle, lung and cortical bone substitutes (CIRS Inc., USA [4]) were used in our measurements for proton range estimation. All samples were manufactured as 10x10 cm slabs of 1, 5, 10 and 20 mm thickness, an additional 20 mm slab was intended as an adapter plate and had a cylindrical cavity for a detector. A difference between nominal and real thickness of the slab stack was less than 0.3% except for 0.6% for the muscle sample.

---

[*] E-mail: vnvasil@orc.ru

The measurements were performed on a horizontal medical proton beam of the ITEP synchrotron facility. The proton beam was spread out by the double scattering technique with a profiled secondary scatterer for better lateral fluence uniformity (see, for example, [5]) and passed through water bellows used as an energy degrader. The beam diameter was limited by a 7 cm steel collimator placed before the energy degrader. A Rogovsky coil was used as a beam monitor, all dose values measured by the detector were normalized to the coil reading to account for the beam instability.

All dose measurements in plastics and liquid water were performed with a silicon diode. Depending on the material under investigation, three experimental setups were used as shown in Fig. 1. Setups A and B were used for the direct CSDA range measurement and subsequent comparison in plastic samples and water. An initial energy of the proton beam was 219 MeV, and a water thickness in the energy degrader was 100 mm. For the plastic measurements (Setup A), the detector was placed in a cavity inside the adapter plate at a distance of 170 mm from the energy degrader window. The investigated plastic slabs have been added in front of the adapter plate by steps from 20 mm at the dose plateau to 1 mm at the Bragg peak.

As the necessary thickness of the lung sample is too high to achieve the Bragg peak due to its low physical density (up to one meter!), it was measured in combination with 13 cm of Plastic Water DT. In that case, the energy degrader to detector distance was 30 cm to place the slabs of both types.

The proton range in liquid water (Setup B) was measured in a water filled PMMA phantom by a 3D scanning device for detector positioning (Fig. 1b). The phantom was set closely to the energy degrader output window; the thickness of the phantom entrance window was 2.5 mm of PMMA and equivalent to 2.9 mm of water, taking into account their linear stopping power ratio. The data was corrected for that thickness in further processing.

As the geometries of measurement in water and plastic (Setup A and B) are different, some correction for a beam divergence and proton fluence change was applied to in-water results. Nevertheless, this correction had only negligible influence, less than 0.1 mm, on the estimated proton range.

Setup C (Fig.1c) was used for estimation of the residual proton range in water after passing 160 mm of the investigated plastic (120 mm for cortical bone). Plastic slabs of that total thickness were placed between the energy degrader output window and the water filled phantom. For comparison, one measurement was performed in water without the plastic, setting the phantom close to the energy degrader. The initial beam energy in these measurements was 220 MeV, the water thickness in the energy degrader was reduced to 20 mm to ensure enough residual range in water.

Finally, depth dose distributions in three types of Plastic Water and cortical bone were calculated by Monte Carlo technique. The proton transport simulation was performed by the program IThMC developed for proton therapy planning and allowing a dose calculation in voxel geometry (up to 512x512x512 voxels). The program takes into account the ionization energy loss in the medium, the energy straggling (by the Landau, Vavilov or normal distributions depending on the material thickness), the elastic multiple coulomb scattering using the Fokker-Planck and Fermi-Eyges models, the elastic and inelastic nuclear reactions on the base of the Sychev model and the D2N2 cross section data set respectively [6].

The simulation geometry was similar to Setup A/B but somewhat simplified. A parallel beam of 200 MeV protons with $\Delta E/E = 0.6\%$ passed through a 100 mm water slab simulating the water energy degrader and a 300 mm slab of the material under investigation. Transport of $10^7$ incident protons was simulated for each material. Depth dose data were calculated along the beam axis using a voxel size of 1 mm and then CSDA ranges were derived from obtained results.

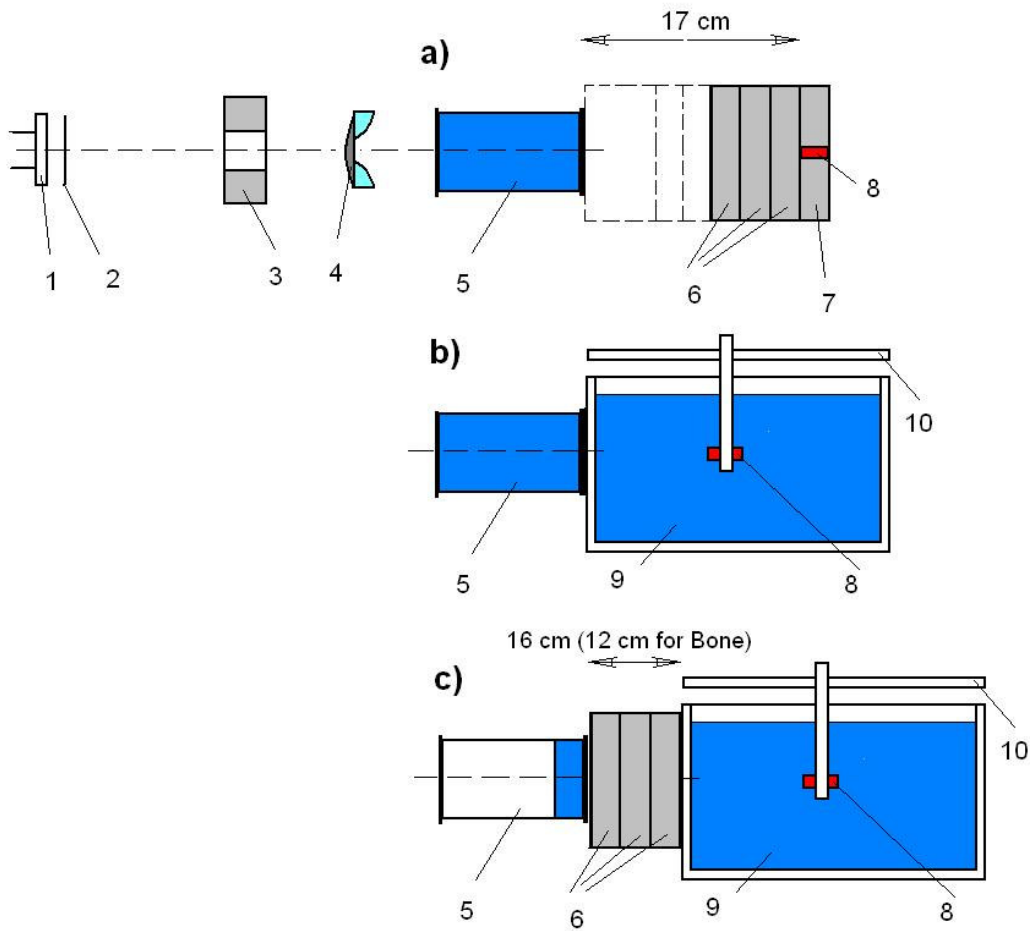

Fig. 1. The measurement setup:
a – direct range measurements in the plastic slabs;
b – direct range measurements in liquid water;
c – residual range measurements in water after the plastics passing.
1 – the proton beam transport system; 2 – the primary scatterer; 3 – the steel collimator; 4 – the secondary profiled scatterer; 5 – the water bellows (energy degrader); 6 – the sample slabs; 7 – the adapter plate; 8 – the diode; 9 – the water filled phantom; 10 – the 3D detector positioner;

## Results

**Direct ranges comparison**

Measured depth dose distributions in all plastics using Setup A and in water using Setup B are shown in Fig. 2. The corrections for proton beam divergence and the phantom wall were applied to the water data as described above. The CSDA range was estimated as the depth distal to the Bragg peak where the dose is reduced to 80% of its maximum value. Mean proton energy at the surface of investigated samples calculated on the base of ICRU49 [3] range-energy relation was estimated to be 153.9 MeV.

All data are normalized to the Bragg peak maximum. Some difference at the dose plateau can be resulted from an uncompensated contribution of the secondary particles generated in the inelastic nuclear reactions.

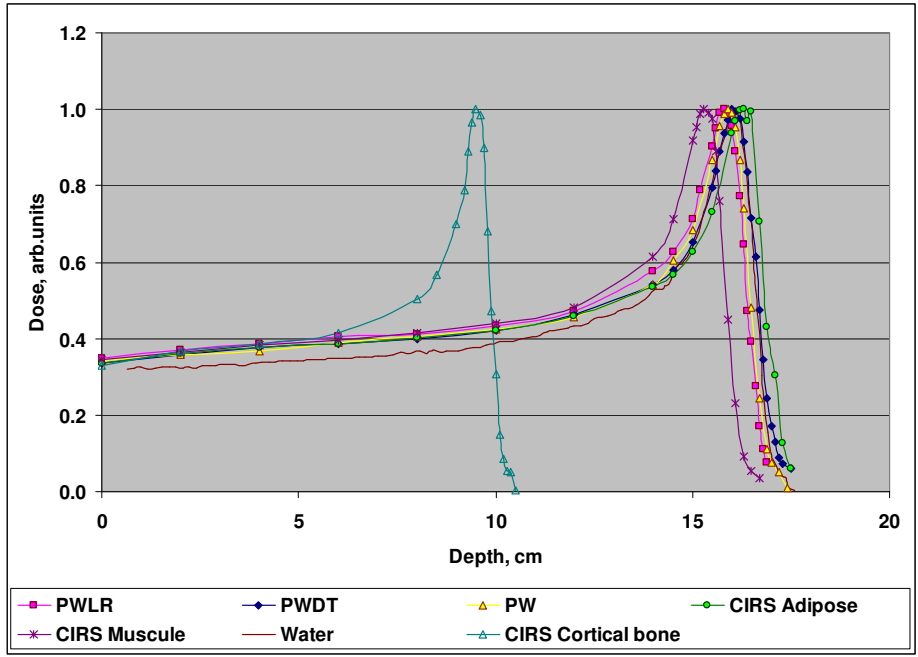

a)

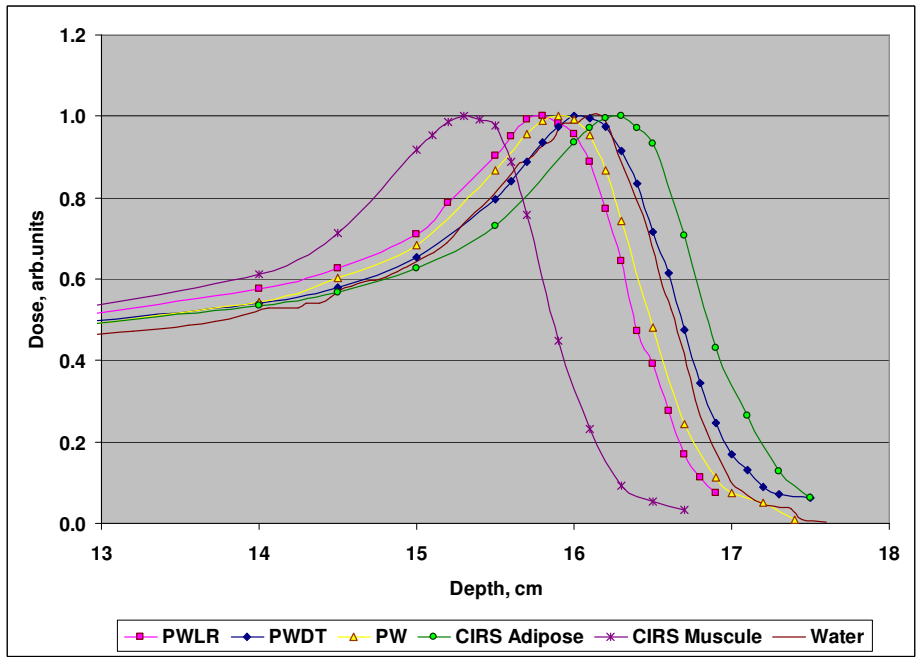

b)

Fig. 2. Measured depth dose distributions in plastics and water: a – the full curves, b – the Bragg peaks.

To estimate the water and tissue equivalence of the investigated materials, the measured proton ranges were compared with respective reference data taken mainly from ICRU Report 49 [3]. In addition, as lung data absent in Report 49, elemental compositions of lung and cortical bone were obtained from the Woodard and White paper [7] and used for proton range calculations with the program SRIM [8, 9].

There is a little discrepancy between these two reference data sets resulted from slightly different excitation potential and the shell correction as discussed in [10, 11]. This discrepancy, as a rule, is within their estimated uncertainties and, for example, is about 1.4% for proton range in water at 150 MeV.

Nevertheless, to avoid the influence of this disagreement, all ranges to compare were related to those in liquid water, experimental or theoretical. As the result, following estimator of the tissue/water equivalence was used:

$$C = \frac{R_m^{meas} / R_w^{meas}}{R_{tissue}^{ref} / R_w^{ref}}, \qquad (1)$$

where $R_m^{meas}$ and $R_w^{meas}$ are measured CSDA proton ranges in the tested substitute and in water; $R_{tissue}^{ref}$ is reference proton range tabulated for the respective tissue (ICRU49 or SRIM); $R_w^{ref}$ is reference proton range in water.

The results of the first measurement series are presented in Table 1.

Table 1. Proton range comparison from direct measurements.

| Substitute | $\rho$, g/cm$^3$ | $R_m^{meas}$, cm | $R_m^{meas} / R_w^{meas}$ | $C$ | Data set | The material for comparison |
|---|---|---|---|---|---|---|
| Plastic Water | 1.030 | 16.22 | 1.013 | **1.013** | ICRU49 | Water |
| Plastic Water LR | 1.029 | 16.19 | 1.010 | **1.010** | ICRU49 | Water |
| Plastic Water DT | 1.039 | 16.47 | 1.028 | **1.028** | ICRU49 | Water |
| Cortical bone | 1.91 | 9.72 | 0.606 | **1.004** | ICRU49 | Cortical bone ICRP, $\rho$ =1.85 g/cm$^3$ |
| Cortical bone | 1.91 | 9.72 | 0.606 | **1.042** | ICRU49 | Compact bone ICRU, $\rho$ =1.85 g/cm$^3$ |
| Cortical bone | 1.91 | 9.72 | 0.592 | **1.024** | SRIM | Woodard-White bone, $\rho$ =1.93 g/cm$^3$ |
| Adipose | 0.96 | 16.60 | 1.036 | **0.981** | ICRU49 | Adipose ICRP, $\rho$ =0.92 g/cm$^3$ |
| Muscle | 1.06 | 15.58 | 0.972 | **1.000** | ICRU49 | ICRP Skeletal Muscle, $\rho$ =1.04 g/cm$^3$ |
| Muscle | 1.06 | 15.58 | 0.972 | **1.002** | ICRU49 | ICRU Striated Muscle, $\rho$ =1.04 g/cm$^3$ |
| Lung +13 cm of PW DT | 0.205 | 16.48 | 4.914 | **0.974** | SRIM | Woodard-White lung, $\rho$ =0.200 g/cm$^3$ |

**Residual range comparison**

The second series of measurements was performed by Setup C (Fig.1c) and allowed to estimate the residual proton range in water after passing 120 (for cortical bone) or 160 mm (for other samples) of the plastic under investigation. Subtracting the obtained residual range from that in water without plastic, an equivalent water thickness was estimated, related to the plastic thickness and then compared with reference range ratio. Thus, in this series of measurements, the tissue/water equivalence estimator $C$ was defined as

$$C = \frac{T_m / (R_w^{meas} - R_{w+m}^{meas})}{R_{tissue}^{ref} / R_w^{ref}}, \qquad (2)$$

where $T_m$ is the thickness of plastic slab, $R_w^{meas}$ and $R_{w+m}^{meas}$ are the measured proton range in water without and with plastic slab, $R_w^{ref}$ and $R_{tissue}^{ref}$ are the reference range in water and in the respective tissue. An advantage of this method is the identity of range measurement conditions for plastic and liquid water.

The obtained results are presented in Table 2.

Table 2. Residual ranges comparison.

| Substitute | $T_m$, cm | $R_{w+m}^{meas}$, cm | $\frac{T_m}{(R_w^{meas} - R_{w+m}^{meas})}$ | $C$ | Data set | The material for comparison |
|---|---|---|---|---|---|---|
| Plastic Water | 15.94 | 7.58 | 0.987 | **0.987** | ICRU49 | Water |
| Plastic Water LR | 15.98 | 7.60 | 0.991 | **0.991** | ICRU49 | Water |
| Plastic Water DT | 16.03 | 7.66 | 0.997 | **0.997** | ICRU49 | Water |
| Cortical bone | 11.93 | 3.54 | 0.591 | **0.981** | ICRU49 | Cortical bone ICRP, $\rho$=1.85 g/cm$^3$ |
| Cortical bone | 11.93 | 3.54 | 0.591 | **1.017** | ICRU49 | Compact bone ICRU, $\rho$=1.85 g/cm$^3$ |
| Cortical bone | 11.93 | 3.54 | 0.591 | **1.000** | SRIM | Woodard-White bone, $\rho$=1.93 g/cm$^3$ |
| Adipose | 15.97 | 8.10 | 1.021 | **0.966** | ICRU49 | Adipose ICRP, $\rho$=0.92 g/cm$^3$ |
| Muscle | 15.89 | 7.00 | 0.950 | **0.978** | ICRU49 | ICRP Skeletal Muscle, $\rho$=1.04 g/cm$^3$ |
| Muscle | 15.89 | 7.00 | 0.950 | **0.980** | ICRU49 | ICRU Striated Muscle, $\rho$=1.04 g/cm$^3$ |
| Lung | 15.95 | 20.36 | 4.738 | **0.941** | SRIM | Woodard-White lung, $\rho$=0.200 g/cm$^3$ |

**Monte Carlo simulation**

As described above, an axial depth dose distribution was calculated in four materials by Monte Carlo technique using the IThMC proton transport code. The geometry of simulation was similar to Setup A/B (Fig.1a and 1b); therefore, the estimator $C$ according to equation (1) was used for the water/tissue equivalence verification. The simulation results are shown in Table 3.

Table 3. Monte Carlo simulation results.

| Substitute | $\rho$, g/cm$^3$ | $R_m^{meas}$, cm | $R_m^{meas}/R_w^{meas}$ | **C** | Data set | The material for comparison |
|---|---|---|---|---|---|---|
| Plastic Water | 1.03 | 15.91 | 1.003 | **1.003** | ICRU49 | Water |
| Plastic Water LR | 1.029 | 15.89 | 1.002 | **1.002** | ICRU49 | Water |
| Plastic Water DT | 1.039 | 15.89 | 1.002 | **1.002** | ICRU49 | Water |
| Cortical bone | 1.91 | 9.49 | 0.598 | **0.991** | ICRU49 | Cortical bone ICRP, $\rho$ =1.85 g/cm$^3$ |
| Cortical bone | 1.91 | 9.49 | 0.598 | **1.028** | ICRU49 | Compact bone ICRU, $\rho$ =1.85 g/cm$^3$ |
| Cortical bone | 1.91 | 9.49 | 0.598 | **1.010** | SRIM | Woodard-White bone, $\rho$ =1.93 g/cm$^3$ |

**Error estimation**

The proton range reproducibility was estimated by repeated measurements in water under the same beam parameters. A range standard deviation was 0.7 mm, it was introduced by the detector positioning accuracy. The plastic slabs thickness and respective detector position in plastic was measured essentially more accurate, the thickness uncertainty never exceeded 0.02 mm. Another significant contribution to the range uncertainty arose from the proton energy instability. A typical energy scatter was 0.5 MeV (i.e. about 0.2%) and resulted in a 1.2 mm proton range error (1SD). Statistical fluctuations of the beam monitor and the diode response resulted in about 0.1 mm of the proton range uncertainty (1SD) calculated by the error propagation formula.

Thus, a total uncertainty of the proton range in water and plastic was estimated to be 1.4 and 1.2 mm (1SD) respectively. That led to an uncertainty of the plastic to water proton range ratio from 1.1 to 1.5% depending on the plastic type in the first measurement series (the direct range comparison) and from 0.9 to 1.3 % in the second measurement series (the residual range comparison). The only exception was the lung equivalent sample in the second series, where the uncertainty reached 5.6%.

## Conclusions

A comparison of Table 1 and 2 demonstrates some systematic difference between the results obtained in direct and residual ranges measurements. Average discrepancy is about 2.5% and can be explained by the measurement uncertainty as estimated above, at least, within ±2SD confidence limits. A systematic nature of this difference allows suggesting the influence of the proton energy instability. A contribution of that factor is to be minimized in future investigations to improve the total accuracy of the results.

The proton ranges in all Plastic Water materials are close to those in liquid water. Theoretical values ratio calculated by Monte Carlo technique is unity within 0.3%. A comparison of the residual ranges in water demonstrates a 0.3-1.2% difference against the reference value. Direct range comparison shows worse agreement – 1.0-1.3% and up to 2.8% for PW DT. The last is the only result lying out of the ±2SD confidence interval.

The range in muscle substitute was compared with that in ICRP skeletal muscle and in ICRU striated muscle. The demonstrated differences were from 0.0-0.2% to 2.0-2.2% in second and first measurement series respectively.

Some proton range underestimation was obtained in the adipose substitute – from 1.9% in the first series to 3.3% in the second one. The last value is situated out of the ±2SD interval.

Conformity of the cortical bone results essentially depends on the reference data used for comparison. ICRU compact bone (ICRU49), ICRP cortical bone (ICRU49) and Woodard-White bone [7] show a 3.6% value scatter. As the result, the proton range discrepancy was 0.4-4.2% and -0.9-+1.7% in the first and second measurement series respectively.

A proton range in lung was underestimated against the Woodard-White/SRIM data in both series – 2.6% in the first and 5.9% in the second one. Nevertheless, the conditions of the second measurement series demonstrated a bad sensitivity in low density materials, like lung, and, respectively, the estimated standard deviation amounting to 5.6%.

Thus, within the estimated uncertainties, Plastic Water, Plastic Water LR, Plastic Water DT, muscle and compact bone demonstrated good agreement with liquid water and respective tissues relative to proton range and stopping power. The range in adipose and lung substitutes is a few percents lower than that in the tissues.